\documentclass[11pt]{article}

\usepackage{fullpage,amsmath,amssymb}
\newcommand{\floor}[1] {\lfloor #1 \rfloor}

\newcommand{\fac} {\mathcal{F}}
\newcommand{\cli} {\mathcal{C}}
\newcommand{\etal}{et al.\ }
\newcommand{\naive}{na\"{\i}ve}
\newcommand{\LP}{\text{LP}}
\newcommand{\OPT}{\text{OPT}}

\newcommand{\I}{\mathcal{I}}

\title{New Results on the Fault-Tolerant Facility Placement Problem}
\author{Li Yan and Marek Chrobak\\
  \{lyan,marek\}@cs.ucr.edu\\
  Computer Science, U of California Riverside\\
  Riverside, CA 92521}
\begin{document}
\maketitle

\section{Introduction}
In the \emph{Fault-Tolerant Facility Placement} problem
(FTFP), we are given a set $\fac$ of \emph{sites} at which
facilities can be built, and a set $\cli$ of \emph{clients}
with some demands that need to be satisfied by different
facilities. A client $j$ has demand $r_j$. Building one
facility at a site $i$ incurs a cost $f_i$, and connecting
one unit of demand from client $j$ to a facility at site
$i\in\fac$ costs $d_{ij}$. The distances $d_{ij}$ form a
metric, that is, they are symmetric and satisfy the triangle
inequality. In a feasible solution, some number of
facilities, possibly zero, are opened at each site $i$, and
demands from each client are connected to those open
facilities, with the constraint that demands from the same
client have to be connected to different facilities. Note
that facilities at the same site are considered different.

It is easy to see that if all $r_j=1$, then FTFP reduces to
the classic uncapacitated facility location problem (UFL).
If we add a constraint that each site can have at most one
facility built, then we get the Fault-Tolerant Facility
Location problem (FTFL). One implication of the one site per
facility restriction in FTFL is that $\max_{j\in\cli} r_j
\leq |\fac|$, while in FTFP $r_j$ can be much bigger than
$|\fac|$.

UFL has a long history and there has been great progress in
designing better approximation algorithms in the past two
decades. Since the publication of the first constant
approximation algorithm by Shmoys, Tardos and
Aardal~\cite{ShmoysTA97}, we have seen a number of
techniques that are applicable in devising approximation
algorithms for UFL with good approximation ratios. Using
LP-rounding, Shmoys, Tardos and Aardal~\cite{ShmoysTA97}
obtained a ratio of 3.16, which was then improved by
Chudak~\cite{ChudakS04} to 1.736, and later by
Sviridenko~\cite{Svi02} to 1.582. Byrka~\cite{ByrkaA10} gave
an improved LP-rounding algorithm, combined with the
dual-fitting algorithm in~\cite{JainMMSV03}, achieving a
ratio of 1.5. Recently Li~\cite{Li11} showed that with a
more refined analysis and randomizing the scaling parameter,
the ratio can be improved to 1.488. This is the best known
approximation result for UFL. Other techniques include Jain
and Vazirani's~\cite{JainV01} primal-dual algorithm with
ratio 3, and Jain, Mahdian, Markakis, Saberi and
Vazirani's~\cite{JainMMSV03} dual-fitting algorithm with
ratio 1.61. Arya \etal\cite{AryaGKMMP04} showed that a local
search heuristic achieves a ratio of 3.

FTFL was first introduced by Jain and
Vazirani~\cite{JainV03} and they adapted their primal-dual
algorithm for UFL to obtain a ratio of $3\ln
(\max_{j\in\cli}r_j)$, which is logarithmic in the maximum
demand. Guha, Meyerson and Monagala~\cite{GuhaMM01} adapted
the Shmoys \etal's~\cite{ShmoysTA97} algorithm for UFL to
obtain the first constant approximation algorithm. Swamy and
Shmoys~\cite{SwamyS08} improved the ratio to $2.076$ using the
idea of pipage rounding. Most recently, Byrka, Srinivasan
and Swamy~\cite{ByrkaSS10} improved the ratio to 1.7245
using dependent rounding and laminar clustering. All the
constant approximation algorithms are based on LP-rounding.

FTFP was first introduced by Xu and Shen~\cite{XuS09} as a
generalization of UFL. They extended the dual-fitting
algorithm~\cite{JainMMSV03} to give an approximation
algorithm with a ratio claimed to be $1.861$. However the
algorithm runs in polynomial time only if $\max_{j\in\cli}
r_j$ is polynomial in $O(|\fac|\cdot |\cli|)$, and the
analysis of the performance guarantee seems flawed.

Our approach is to reduce FTFP to FTFL. We then use Byrka
\etal's~\cite{ByrkaSS10} result on FTFL to achieve the ratio
$1.7245$.

\section{The LP Formulation}
The FTFP problem has a natural IP formulation. Let $y_i$
represent the number of facilities built at site $i$ and let
$x_{ij}$ represent the number of connections from client $j$
to facilities at site $i$. If we relax the integral
constraints, we obtain the following LP:

\begin{alignat}{3}
  &\textrm{minimize} &\quad&\textstyle{\sum_{i\in
      \fac}f_iy_i + \sum_{i\in \fac, j\in \cli}d_{ij}x_{ij}}
  \label{eqn:fac_primal}\\ \notag
  &\textrm{subject to}&& y_i - x_{ij} \geq 0, &\quad& \forall i\in \fac,
  j\in \cli \\ \notag
  & &&\textstyle{
    \sum_{i\in \fac} x_{ij} \geq r_j,} && \forall j\in \cli
  \\ \notag
  & &&\textstyle{
   x_{ij} \geq 0, y_i \geq 0,} && \forall i\in \fac, j\in \cli 
  \\ \notag
\end{alignat}

The dual program is:
\begin{alignat}{3}
  &\textrm{maximize}&\quad& \textstyle{\sum_{j\in \cli} r_j\alpha_j}
  \label{eqn:fac_dual}  \\ \notag
  &\textrm{subject to} && \textstyle{
    \sum_{j\in \cli}\beta_{ij} \leq f_i,}  &\quad& \forall i \in \fac  \\ \notag
  & &&\textstyle{\alpha_{j} - \beta_{ij} \leq
    d_{ij},}  && \forall i\in \fac, j\in \cli \\ \notag
  & &&\textstyle{\alpha_j \geq 0,
    \beta_{ij} \geq 0,} && \forall i\in \fac, j\in \cli
  \\ \notag
\end{alignat}

\section{A 1.7245-Approximation Algorithm}
In this section we give an algorithm based on a reduction to
FTFL. Moreover, we show that if we have a
$\rho$-approximation algorithm for FTFL, then we can use it
to build an approximation algorithm for FTFP with the same
ratio. To be more precise, we assume that $\rho$ is the
ratio of the cost of the FTFL solution to the fractional
optimal solution cost of the natural LP formulation for
FTFL.~\footnote{Adding a constraint $y_i\leq 1,\ \forall
  i\in\fac$ to the LP (\ref{eqn:fac_primal}) results in the
  LP for FTFL.}

A \naive\ idea is to split the sites so that each site is
split into $P=\max_{j\in\cli}r_j$ identical sites. Now we
can restrict each split site to have at most one
facility. Since we never need more than $P$ facilities open
at the same site in the original instance, the FTFL instance
after split is equivalent to the original FTFP
instance. However, since $P$ might be large, this reduction
might result in an instance with exponential size.

However, we can modify the \naive\ approach to obtain an FTFL
instance with polynomial size, using an optimal fractional
solution to the LP~(\ref{eqn:fac_primal}).

Let $(\mathbf{x}^\ast,\mathbf{y}^\ast)$ be the fractional
optimal solution to LP~(\ref{eqn:fac_primal}). Also let
\begin{equation*}
  \hat y_i = \max\{0,\floor {y_i^\ast} -1\},\ \hat x_{ij}  =
  \min\{\floor {x_{ij}^\ast},\hat y_i\},\ 
\end{equation*} and
\begin{equation*}
  \bar x_{ij} = x_{ij}^\ast - \hat x_{ij},\ \bar y_i =
  y_i^\ast - \hat y_i.  
\end{equation*}
Let $\I$ be the original FTFP instance. We define an FTFP
instance $\I_1$ with the same set $\fac$ of sites and $\cli$
of clients, the same distances $d_{ij}$ and the same
facility costs $f_i$, except that the demands are $\hat
r_j=\sum_{i\in\fac}\hat x_{ij}$. Let $S_1$ be the solution
defined by $(\hat x_{ij}, \hat y_i)$. Clearly $S_1$ is a
feasible integral solution to the instance $\I_1$.

Another FTFP instance $\I_2$ is defined using the same
parameters except that the demands are $\bar r_j= r_j-\hat
r_j=\sum_{i\in\fac}\bar x_{ij}$. We claim that $(\bar
x_{ij},\bar y_i)$ is a feasible fractional solution to the
instance $\I_2$~\footnote{Note that if we take $\hat
  x_{ij}=\floor{x^\ast_{ij}}, \hat y_i=\floor{y^\ast_i}$ and
  define $\bar x_{ij},\bar y_i, \bar r_j$ in a similar way,
  then $(\bar x_{ij}, \bar y_i)$ may not be feasible to the
  instance with demands $\{\bar r_j\}$ because we might have
  $\bar x_{ij} > \bar y_i$ if $0<x^\ast_{ij}<1$ and
  $y^\ast_i=1$.}. To ensure feasibility, we only need to
show $\bar x_{ij} \leq \bar y_i$ for all
$i\in\fac,j\in\cli$. The proof proceeds by considering two
cases. Case 1 is that $y^\ast_i<1$. Then we have $\bar
x_{ij} = x^\ast_{ij}\leq y^\ast_{i}=\bar y_i$ as
needed. Case 2 is $y^\ast_i\geq 1$. From the definition we
have $\bar y_i \geq 1$. We further consider two subcases: if
$\hat x_{ij}=\hat y_i$, then $\bar x_{ij}=x^\ast_{ij}-\hat
x_{ij} \leq y^\ast_i-\hat y_i\leq\bar y_i$ and we are
done. The other subcase is $\hat x_{ij} < \hat y_i$. It
follows that $0\leq \bar x_{ij} < 1$. Again we have $\bar
x_{ij}<1\leq \bar y_i$. The above takes care of all cases,
so we have shown the feasibility of the fractional solution
$(\bar x_{ij}, \bar y_i)$ to the instance $\I_2$.

One more observation about the instance $\I_2$ is that $\bar
r_j \leq 2n$ where $n=|\fac|$, since $\bar r_j =
\sum_{i\in\fac}\bar x_{ij}$ and for every
$i\in\fac,j\in\cli$, we have $0\leq\bar x_{ij}<2$. Therefore
we have $\bar P=\max_{j\in\cli} \bar r_j \leq 2n$. Now that
we have the demands being polynomial w.r.t.\ the input size
of the original FTFP instance, we can split each site into
$\bar P$ new sites and treat the instance as an FTFL
instance. Solving the FTFL instance with a
$\rho$-approximation algorithm, we have an integral solution
$S_2$ with cost no more than $\rho\cdot \LP^\ast(\I_2)$,
where $\LP^\ast(\I_2)$ is the cost of a fractional optimal
solution to the instance $\I_2$.

Combining $S_1$ and $S_2$, we have a feasible solution to
the original FTFP instance. We now argue that the solution
is within a factor of $\rho$ from the cost of an optimal
fractional solution to the LP~(\ref{eqn:fac_primal}).

First we observe that $cost(S_1)=\sum_{i\in\fac}f_i\hat y_i
+ \sum_{i\in\fac,j\in\cli}d_{ij}\hat x_{ij}$. Secondly,
since $(\bar x_{ij},\bar y_i)$ is a feasible fractional
solution to the instance $\I_2$, we have
$\sum_{i\in\fac}f_i\bar y_i + \sum_{i\in\fac,j\in\cli}
d_{ij}\bar x_{ij} \geq \LP^\ast(\I_2)$. Our solution has
total cost
\begin{align*}
  cost(S_1)+cost(S_2) &\leq \sum_{i\in\fac}f_i\hat y_i +
  \sum_{i\in\fac,j\in\cli} d_{ij}\hat x_{ij} + \rho\cdot
  \LP^\ast(\I_2)\\
  &\leq \sum_{i\in\fac}f_i\hat y_i +
  \sum_{i\in\fac,j\in\cli} d_{ij}\hat x_{ij} +
  \rho\ (\sum_{i\in\fac}f_i\bar y_i + \sum_{i\in\fac,j\in\cli}
  d_{ij}\bar x_{ij})\\
  &\leq \rho\ (\sum_{i\in\fac}f_i\hat y_i +
  \sum_{i\in\fac,j\in\cli} d_{ij}\hat x_{ij} +
  \sum_{i\in\fac}f_i\bar y_i + \sum_{i\in\fac,j\in\cli}
  d_{ij}\bar x_{ij})\\
  &= \rho\ (\sum_{i\in\fac} f_i y_i^\ast + d_{ij}
  x_{ij}^\ast) = \rho\cdot \LP^\ast(\I)\\
  &\leq \rho\cdot \OPT(\I)\\
\end{align*}

The currently best known approximation algorithm for FTFL
achieves a ratio of $1.7245$~\cite{ByrkaSS10}, so our
algorithm is a $1.7245$-approximation algorithm for FTFP.

\section{Approximation for Large Demands}
If we seek an approximation algorithm with ratio as a
function of $R=\min_{j\in\cli} r_j$, then we show that we
can achieve a ratio of $1+O(n/R)$ where $n=|\fac|$. The
hidden constant in the big-O is the same as the best
approximation ratio for FTFL, which is known to be at most
$1.7245$~\cite{ByrkaSS10}.~\footnote{The hidden constant can
  be improved to be $1.488$ by solving a sequence of UFL
  instances.}

Our algorithm uses the $1.7245$-approximation algorithm for
FTFL as a subroutine. First we solve the LP
(\ref{eqn:fac_primal}) and obtain $(x_{ij}^\ast,y_i^\ast)$
as the optimal fractional solution. Now we round down the
fractional values as before and the rounded down values
$\hat x_{ij} = \floor{x_{ij}^\ast}, \hat y_i =
\floor{y_i^\ast}$ form part of the final solution, denoted
as $S_1$. Now each client $j$ has a reduced demand $\bar r_j
= r_j - \sum_{i\in\fac} \hat x_{ij} \leq n-1$, where
$n=|\fac|$. Therefore we can solve this instance with
reduced demands, denoted as $\I_2$, by solving an equivalent
FTFL instance: We replicate each site with $n-1$ copies
since no site needs to have more than $n-1$ facilities
open. Now we can restrict each duplicated site open at most
one facility and this gives us an FTFL instance with
polynomial size to work with. We then use the
$1.7245$-approximation algorithm to solve this FTFL instance
to obtain an integral solution $S_2$ with
$cost(S_2)\leq\rho\cdot\LP^\ast(\I_2)$ with
$\rho=1.7245$. We now show that $\LP^\ast(\I_2) \leq
\frac{n}{R}\LP^\ast(\I)\leq \frac{n}{R}\OPT(\I)$, where $\I$
is the input FTFP instance. To see this, consider three FTFP
instances: $\I_1$ is the instance with all $r_j=R$, and
$\I_2$ is the instance with all $r_j=\bar{r}_j$, and $\I_3$
is the instance with all demands $r_j=n-1$. We have
$\LP^\ast(\I_3)/\LP^\ast(\I_1)=(n-1)/R$ since any fractional
solution with cost $Z$ of an FTFP instance with uniform
demands $s$ can be scaled down by $s$ to obtain a feasible
solution for the same instance with demands set to $1$ and
the cost being $Z/s$. We also have $\LP^\ast(\I_3)\geq
\LP^\ast(\I_2)$ since every demand in $\I_3$ is at least as
large as that in $\I_2$. Similarly $\LP^\ast(\I) \geq
\LP^\ast(\I_1)$. Therefore we have
\begin{equation*}
  \frac{\LP^\ast(\I_2)}{\LP^\ast(\I)} \leq
  \frac{\LP^\ast(\I_3)}{\LP^\ast(\I_1)} = \frac{n-1}{R} \leq \frac{n}{R}.
\end{equation*}

By combining $S_1$ and $S_2$ we have a feasible solution to
the original FTFP instance. Now we bound the total
cost. First notice that $cost(S_1)\leq \LP^\ast(\I) \leq
\OPT(\I)$. By the previous argument, we have $cost(S_2)\leq
\rho \frac{n}{R} \OPT(\I)$. Therefore, the total cost of our
solution is
\begin{equation*}
  cost(S_1)+cost(S_2) \leq (1+\rho\frac{n}{R}) \OPT(\I),
\end{equation*}
which is $(1+O(\frac{n}{R}))\OPT(\I)$. The approximation
ratio approaches $1$ when $R$ becomes much larger than
$n$. This is perhaps not too surprising as when all the
demands are large, the fractional optimal solution to the LP
(\ref{eqn:fac_primal}) is close to integral relative to the
demands, so their costs are close.

\bibliographystyle{plain}
\bibliography{facility}

\end{document}